\documentclass[10pt]{iopart}
\usepackage{iopams}
\usepackage{graphicx}

\begin{document}

\title[Using ion production to monitor the birth and decay]
{Using ion production to monitor the birth and death of a
metastable helium Bose-Einstein condensate}

\author{S. Seidelin\footnote[3]{To whom correspondence
should be addressed (signe.seidelin@iota.u-psud.fr)}, O. Sirjean,
J. Viana Gomes\footnote[1]{Permanent address: Departamento de
Fisica, Universidade do Minho, Campus de Gualtar, 4710-057 Braga,
Portugal}, D. Boiron, C. I. Westbrook and A. Aspect  }

\address{Laboratoire Charles Fabry de l'Institut
d'Optique, UMR 8501 du CNRS, F-91403 Orsay Cedex, France}

\begin{abstract}
We discuss observations of the ion flux from a cloud of trapped
$2~^{3}\textrm{S}_{1}$ metastable helium atoms. Both Bose-Einstein
condensates and thermal clouds were investigated. The ion flux is
compared to time-of-flight observations of the expanded cloud. We
show data concerning BEC formation and decay, as well as
measurements of two- and three-body ionization rate constants. We
also discuss possible improvements and extensions of our results.
\end{abstract}

\maketitle

\section{Introduction}
Metastable helium (He*), has recently joined the list of atomic
species for which Bose-Einstein condensates (BEC) have been
realized \cite{BecHe,BecENS}. Its major particularity is the 20 eV
internal energy of the metastable state. Although this
metastability leads to additional possible loss channels, it has
been shown that these are not a serious problem. Indeed, ionizing
collisions are a benefit because their low rate is nevertheless
easily detectable. Ion detection is thus a  new,
``non-destructiv'' and real-time observation tool for studies of
phenomenon of BEC formation kinetics
\cite{NaissanceBEC,kohl,yum,gar,stoof}. In this paper we will
describe our progress toward rendering quantitative the ion
signal.

Several loss mechanisms are specific to the metastable state.
First, collisions with the background gas lead to Penning
ionization of the background gas:
$$X + He^* \rightarrow X^+ +He + e^-$$
The positive ion $X^+$ produced can be easily detected and if this
is the dominant ion production mechanism as it is for a dilute
sample (for a density  $n \lesssim 10^{10}\;{\rm{cm^{-3}}} $), the
corresponding flux is proportional to the number of trapped He$^*$
atoms. So for example we can easily measure the lifetime of a
dilute, trapped sample. This linearity no longer holds when the
density of the trapped cloud becomes high. Collisions between
atoms in the cloud must be taken into account. The relevant
ionization mechanisms involve both two-body processes:
\begin{equation}\label{2corps}
He^{*} + He^{*} \rightarrow \left \{
\begin{array}{l}
He^{+} + He(1S) + e^{-}\\
He_{2}^{+} + e^{-}
\end{array}
\right.
\end{equation}
as well as a three-body process:
\begin{equation}\label{3corps}
\begin{array}{rccl}
He^{*} + He^{*} + He^{*}& \rightarrow & He_{2}^{*}& + He^{*} (\sim
1mK) \\
 & & \hookrightarrow & He^{+} + He(1S) + e^{-}
\end{array}
\end{equation}
When these processes are present, the ion flux is related to the
spatial integral of $n^2$ and $n^3$. At BEC densities, the 2- and
3-body processes dominate the background gas ionization, and so
detecting the ion flux in this case amounts
to monitoring the atomic density.\\

In this paper, after a rapid description of our experimental
setup, we present observations, via the ion flux, of the formation
and the decay of a He* BEC. The observations are mainly
qualitative, but we discuss some of the requirements for making
them quantitative. We then discuss our measurements of the 2- and
3-body ionization rate constants both in a BEC \cite{prl} and in a
thermal cloud. We discuss some of the systematic errors in these
measurements and conclude with some ideas for avoiding these
errors.

\section{Setup and experimental procedure}

Our setup has been described previously
\cite{BecHe,prl,Theseantoine}. Briefly, we trap up to $2\times
10^8$ atoms at 1~mK in a Ioffe-Pritchard trap with a lifetime
($\tau$) of 90~s. We use a ``cloverleaf" configuration
(Fig.\ref{fig1})  \cite{Mewes:96} with a bias field $B_0 =300$~mG.
The axial and radial oscillation frequencies in the harmonic
trapping potential are typically $\nu_{\parallel}=47\pm3$~Hz and
$\nu_{\bot}=1200\pm50$~Hz respectively
($\overline{\omega}/2\pi=(\nu_{\parallel}\nu_{\perp}^{2})^{1/3}=408$~Hz).
In a typical run, forced evaporative cooling takes place for 40~s
and is divided into 4 linear ramps. The last ramp lasts 5 seconds
and the frequency decreases from 2000~kHz to a value between 1500
and 1000~kHz, depending on the condensed fraction wanted. A
frequency of 1000~kHz (which is about 50~kHz above the minimum of
the trapping potential) corresponds to the formation of a pure
condensate.

A special feature of our set up is the detection scheme, based on
a 2 stage, single anode microchannel plate detector (MCP) placed 5
cm below the trapping region (Fig.\ref{fig1}). Two grids above the
MCP allow us either to repel positive ions and detect only the He*
atoms, or to attract and detect positive ions produced in the
trapped cloud.
\begin{figure}
\begin{center}
\includegraphics[height=8cm]{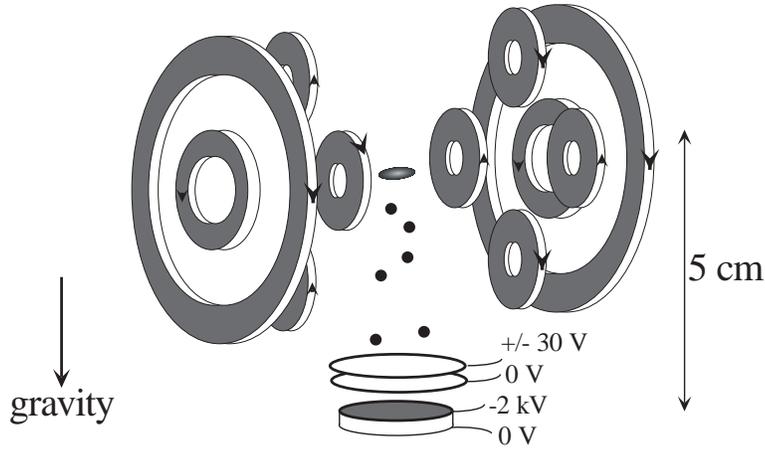}
\end{center}
\caption{\textit{Experimental setup. The cold atoms are trapped in
a cloverleaf type magnetic trap. A special feature of our set up
is the microchannel plate detector (MCP) placed below the trapping
region. Two grids above the MCP allow us either to repel positive
ions and detect only the He* atoms suddenly released from the trap
(time-of-flight measurements), or to attract and detect the
positive ions produced in the trapped cloud (ion rate
measurements). }}\label{fig1}
\end{figure}
To detect the ion flux, the MCP is used in counting mode
\cite{prl}: the anode pulses from each ion are amplified, and
processed by a counter which records the time delay between
successive events. We can also use the MCP to record a
time-of-flight signal (TOF) of the atoms released from the trap.
Because the width of the TOF distribution is small (about 5 ms for
a BEC) compared to the mean arrival time (100 ms), all of the
atoms hit the detector with nearly the same final velocity of 1
m/s. The TOF spectra are then proportional to the spatial
distribution along the vertical direction, integrated over the two
horizontal directions. To record the TOF we use the MCP in analog
mode to avoid saturation due to a very high instantaneous flux
\cite{prl}.

\section{Monitoring the evolution of a He$^*$ cloud}\label{monitor}

To monitor the evolution of an atomic cloud, one usually releases
the cloud and measures the TOF signal. Such a technique is
destructive, and one must repeat the cooling sequence for each
measurement. The TOF signals are thus subject to fluctuations in
the initial number of atoms. In our case, we have a supplementary
signal: the ion rate. We can thus minimize these fluctuations, by
selecting runs having identical ion rates from the time between
the beginning of the last rf-ramp until release.

Another type of observation is possible, however. We can use the
evolution of the {\it value} of the ion rate, which is obtained in
a single run, independent of any initial fluctuations. When the
density is close to the density for BEC formation (i.e. $n \gtrsim
10^{12}\;{\rm{cm^{-3}}}$), 2- and 3-body collisions within the
cloud dominate the ion production. Thus the ion rate is related to
the density of the cloud via the 2- and 3-body rate constant.
Under some conditions (see \ref{ions}) a record of the ion rate
followed by a TOF measurement at the end of the formation of the
BEC allows one to monitor the evolution of all the parameters of
the cloud. In such an observation, knowledge of the 2- and 3-body
rate constants is essential. This is the aim of the experiments
described in section \ref{ratecst}.

\subsection{Observation of condensate formation during the
evaporation ramp}

\begin{figure}
\begin{center}
\includegraphics[height=6cm]{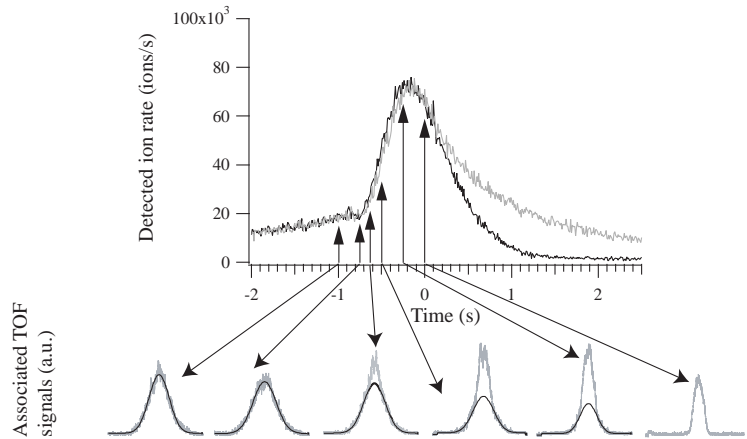}
\end{center}
\caption{\textit{Single shot measurements of the ion rate versus
time and the corresponding TOF signals. Forced evaporative cooling
takes place until $t=0$ (only the last 2 seconds of the RF-ramp
are shown : from 1400~kHz to 1000~kHz). The arrows indicate the
time the trap was switched off to record the TOF. The black curves
superimposed on the TOF signals are Gaussian fits to the wings of
the TOF. See also caption of Fig. \ref{VieBec}.}}\label{naiss}
\end{figure}

Before trying to do a quantitative experiment on BEC formation out
of a non-equilibrium uncondensed cloud \cite{NaissanceBEC,kohl},
we can explore qualitatively what happens during our standard
evaporation ramp. We show in Fig. \ref{naiss} the evolution of the
ion rate from $2$ seconds before the end of the rf-ramp to $2.5$
seconds after it. In addition we show the TOF signals
corresponding to various times before the end of the ramp,
selected using their initial ion rate. Between times $t=-2$ s and
$t=0$, the rf-frequency was ramped down linearly from $1.4$ MHz to
$1$ MHz. At $t=0$ a pure condensate is formed. The comparison of
the TOF and ion data first shows that the appearance of a narrow
structure in the TOF spectrum corresponds, as closely as we can
observe it, to an abrupt change in the slope of the ion signal.
Thus, not only is the ion signal a reliable indicator of the
presence of a BEC, but also a precise measure of the time of its
appearance.

One's first reaction in looking at the ion rate signal is to
assume that the higher the ion signal, the larger the BEC and the
smaller the thermal cloud. Fig. \ref{naiss} shows however, that
this is not quite the case: the maximum in the ion signal arrives
before the achievement of a pure BEC. In fact, computing the value
of the ion signal is rather complex. First, as was discussed in
Refs. \cite{DepletionGora,Burt}, as well as below, the
indistinguishability of the atoms in the BEC renders the effective
2- and 3-body collision rate constants smaller than in the thermal
cloud by factors of $1/2!$ and $1/3!$, respectively. Collisions
between condensed and non-condensed atoms must also be taken into
account \cite{DepletionGora} and the degree of overlap between the
two clouds must be calculated. Thus it might be conceivable to see
the ion rate go down when a BEC is formed. We show however in
\ref{ions} that for a fixed total number of atoms, the ion rate
monotonically increases as a BEC becomes more and more pure. The
observation in Fig. \ref{naiss} is explained by the fact that, up
until $t=0$ in Fig. \ref{naiss}, the atoms are being evaporatively
cooled, as well as undergoing ionizing collisions and thus the
total number of atoms must be decreasing. An explicit calculation
including the atom loss is given in \ref{ions} and agrees
qualitatively with our observations.

\subsection{Observing the decay of the condensate}

\begin{figure}
\begin{center}
\includegraphics[height=6cm]{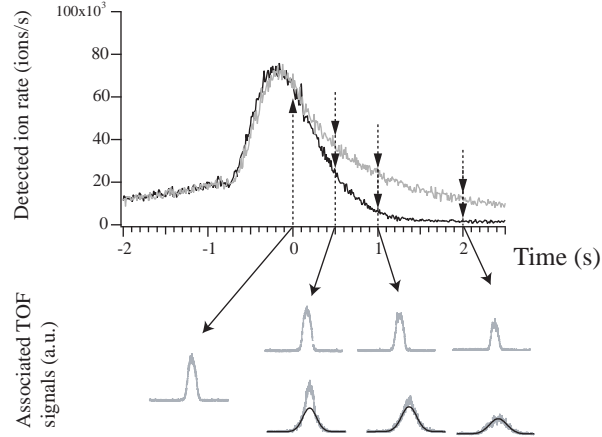}
\end{center}
\caption{\textit{Same as the previous figure except that we
examine the decay of the ion signal after $t=0$. The upper TOF
curves correspond to the upper (grey) ion decay curve (rf-shield
present). The lower TOF curves correspond to the lower black ion
curve (without rf-shield). This shows that the rf-shield is
maintaining a quasi-pure BEC during the decay, and that in the
absence of an rf-shield the condensate rapidly heats up, causing
the ion rate to drop even faster.}}\label{VieBec}
\end{figure}

\begin{figure}
\begin{center}
\includegraphics[height=4cm]{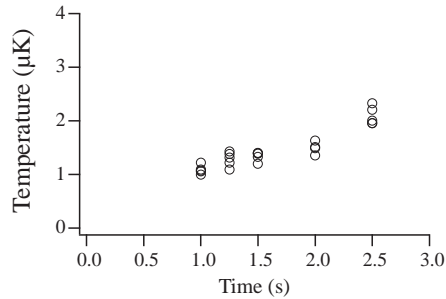}
\end{center}
\caption{\textit{Heating of the condensate in the absence of an
rf-shield. The temperature increases from 1.1 $\mu K$ to 2.2 $\mu
K$ in 1.5 seconds. The time $t=0$ is the same as in Fig.
\ref{naiss} and \ref{VieBec}. For each different time, 4 different
TOF's have been acquired and fitted.}}\label{heating}
\end{figure}

Figure \ref{VieBec} shows a series of TOF spectra taken after the
end of the rf-ramp. Two situations are shown. In one case (grey
ion curve) the rf-knife was held on at the frequency corresponding
to the end of the ramp.  In the other case (black ion curve) the
rf-power was turned off completely at the end of the ramp. The
data show that the condensate remains pure with the rf-knife kept
on. In the absence of the rf-knife, the ion rate decays much
faster and one sees that the sample rapidly acquires a thermal
component. Since the total number of trapped atoms in the presence
of a knife must be smaller than or equal to that in the absence of
rf-knife, we conclude that the rapid decline in ion rate is due to
a loss in sample density and not in the total number of atoms.
This conclusion is confirmed by a fit to the thermal wings which
reveals a heating as shown in Fig. \ref{heating}.

\subsection{Measuring the total number of atoms.}

\begin{figure}
\begin{center}
\includegraphics[height=4cm]{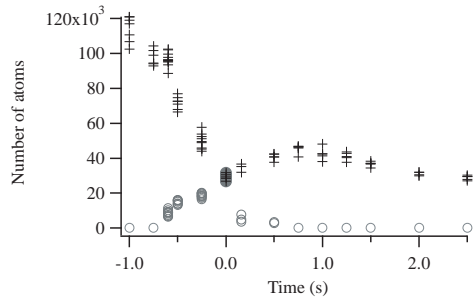}
\end{center}
\caption{\textit{The measured number of atoms as a function of
time. Crosses represent the total atom number, circles represent
the number of atoms in the condensed part. The data comes from the
fits of the TOF's presented in Fig. \ref{naiss} and \ref{VieBec}
and corresponds to the case where the rf knife is absent. The time
scale indicated is the same as in Fig. \ref{naiss} and
\ref{VieBec}. The increase in the total number after $t=0$ is
spurious (see text).}}\label{Natomes}
\end{figure}

An attempt to measure the total number of atoms as a function of
time is shown in Fig. \ref{Natomes}. Both the total number and the
condensed number as derived from fits to the TOF signals of Fig.
\ref{naiss} and Fig. \ref{VieBec} are plotted. Surprisingly the
total number of atoms appears to increase between $t=0$ and
$t=1$~s. There must be a systematic error, which we can account
for by recalling that in our apparatus we only detect atoms which
make non-adiabatic transitions to the (field insensitive) $m=0$
state during the turn-off of the magnetic trap \cite{BecHe}. The
fraction we observe is of order 10\%. It is quite possible that
this non-adiabatic transition does not occur with equal
probability at every point in the trap. Thus clouds with different
spatial distributions may be converted to the $m=0$ state with
different efficiencies. This could explain why atoms in the
thermal cloud are observed with a higher efficiency than condensed
atoms, as indicated in Fig. \ref{Natomes}.

We conclude that our measurements of the absolute number of atoms
contain uncontrolled systematic errors of the order of a factor of
two. So, even if we knew the ionization rate constants, we cannot
use the ion rate to study condensate growth kinetics because we
need the absolute value of the initial number of atoms and it
would also be useful to measure the variation of the number of
atoms during formation. Such a study will have to wait for a more
reliable method of releasing the atoms from the trap (see
conclusion). However, the measurement of the ionization rate
constants is a first step. For a BEC, we can circumvent the
systematic error on the detection efficiency of the atoms to make
a measurement of the ionization rate constants. This has been
described in Ref. \cite{prl} and will be summarized in the
following section. Afterwards, we will investigate the effect of
this systematic error on the measurement for a thermal cloud.

\section{\bf Rate constants of ionizing collisions}\label{ratecst}

The usual method of measuring the inelastic rate constants relies
on fits to a non-exponential decay of the number of atoms. This
method has some practical problems if the sample heats during the
measurement: the density changes and complicates the fitting
procedure. A way to avoid this heating is to apply an rf-shield,
but this latter causes atom losses, which are not due to
collisions. What is even more inconvenient in our case, is that
what is measured in this kind of experiment is a decreasing atom
number due to losses, which can be due to ionizing as well as
non-ionizing collisions. We want to relate the ion rate to the
density of the cloud, so what we need is the rate constants for
{\it{ionizing}} 2- and 3-body collisions. We therefore use another
method which consists in directly observing the products of the
collisions, namely the number of ions, as a function of the
density of the cloud.

As we have seen in section \ref{monitor}, there is a systematic
error on the measurement of the number of atom and so of the
density of the cloud. But we will see that we can circumvent it in
the case of the BEC. Let us then assume in a first part that we
are able to measure accurately the number of atoms.

We use the MCP to detect both the ions and the TOF signal. In a
single run we record the ion rate during the last seconds of the
ramp until we switch off the magnetic trap and record the TOF
signal (to obtain the atom number $N$ and the density). The very
last value of the ion rate recorded corresponds to ions produced
by the cloud observed with the TOF signal. We repeat this sequence
many times with different numbers of atoms in the cloud. The way
to vary this number is to keep the atoms in the trap with an
rf-shield kept on. In this way we reduce the atom number and keep
the temperature of the cloud constant. As explained in \ref{ions},
the relation between ion rate and density is quite complex in the
case of the presence of collisions between atoms in the condensed
part and atoms in the thermal part. We therefore only examine the
case of a pure BEC {\it{or}} a pure thermal cloud. In that case we
can write the ion rate per atom $\Gamma$ as follows

\begin{equation}\label{formulegamma}
\frac{{\rm{ion~ rate}}}{N}=\Gamma= \frac{1}{\tau'} + \frac{1}{2}
\; \kappa_{2} \; \beta \; \langle n \rangle + \frac{1}{3} \;
\kappa_{3} \; L \; \langle n^{2} \rangle.
\end{equation}
where $\langle n \rangle=\frac{1}{N}\int{n^2~d{\bf{r}}}$ and
$\langle n^2 \rangle=\frac{1}{N}\int{n^3~d{\bf{r}}}$, $n$ being
the local density. We have also introduced the 2-body and 3-body
ionizing collision rate constants, $\beta$ and $L$, respectively,
defined according to their effect on the density loss in a thermal
gas \cite{notedefcoef}: $ (\frac{dn}{dt})_{ionization} = -
\frac{n}{\tau'} - \beta \, n^{2} - L \, n^{3}$. The effective
lifetime $\tau' \geq \tau$ is due to {\it{ionizing}} collisions
with the background gas. The numerical factors come from the fact
that although 2 or 3 atoms are lost in each type of collision,
only 1 ion is produced. The factors $\kappa_{2}$ and $\kappa_{3}$
take into account the fact that the 2 and 3-particle local
correlation functions are different depending on whether it is a
BEC or a thermal cloud. For the thermal cloud
$\kappa_{2}=\kappa_{3}=1$, while for a dilute BEC, one has
$\kappa_{2}=1/2!$ and $\kappa_{3}=1/3!$ \cite{DepletionGora,Burt}.
When the sample is very dense, quantum depletion must be taken
into account, which modifies these factors \cite{DepletionGora}. A
measurement of $\beta$ and $L$ would allow us to test
experimentally the theoretical values of $\kappa_{2}$ and
$\kappa_{3}$ \cite{prl}.

\subsection{\bf Rate constants for a BEC}\label{cstbec}

To determine the ionizing collision rate constants $\beta$ and
$L$, we need an absolute calibration of the number of atoms in the
condensate $N_{0}$, and the peak density $n_{0}$ in order to
calculate $\langle n \rangle$ and $\langle n^2 \rangle$. As
discussed above, we do not have a good calibration of these
quantities. In the case of a BEC however, the measurement of the
chemical potential $\mu$ obtained by a fit of the TOF signal,
gives an accurate value for the product $n_{0} \, a =\mu m
/4\pi\hbar^{2}$, $a$ being the scattering length. With the value
of $\overline{\omega}$ we also obtain the product $N_{0} \, a
=(1/15) \, (\hbar / m \overline{\omega})^{1/2}\, (2\mu /
\hbar\overline{\omega})^{5/2} $. Experimentally we confirm that
$\mu \propto N_{d}^{2/5}$ where $N_{d}$ is the number of detected
atoms \cite{prl}. This is a good indication that our detector is
linear and that the detection efficiency for a BEC is indeed
independent of $\mu$. Assuming a value of the scattering length
($a=20 \, \textrm{nm}$), we have therefore an accurate measurement
of $n_0$ and $N_0$. We have measured the rate constants $\beta$
and $L$ for a condensate \cite{prl}. We obtain by a fit to
equation (\ref{formulegamma}) (having corrected for the effect of
quantum depletion and the fact that the BEC also contains a small
thermal fraction) $ \beta = 2.9 (\pm2.0) \times 10^{-14}$~cm$^{3}$
sec$^{-1}$ and $L = 1.2 (\pm 0.7) \times 10^{-26}$~cm$^{6}$
sec$^{-1}$. These values agree with the theoretical estimates
\cite{Shlyapnikov:94a,venturi:99}. The scattering length is not
well-known \cite{BecHe,BecENS}, so we have also given $\beta$ and
$L$ for different values of $a$ \cite{prl}.

\subsection{\bf Rate constants for a thermal cloud}

To determine the rate constants for a thermal cloud we need, as
before, to determine the atom number and density. We cannot use
the same trick as in section \ref{cstbec} to avoid systematic
errors in the detection efficiency. If we want to use the above
experimental method for a thermal cloud we must rely on a fit of
the TOF to find the atom number and the temperature $T$. In
\ref{detection}, we propose a method to determine the rate
constants which is independent of an absolute detection
efficiency, but at this stage we will concentrate on the same
technique as used for a BEC.

\begin{figure}
\begin{center}
\includegraphics[height=4cm]{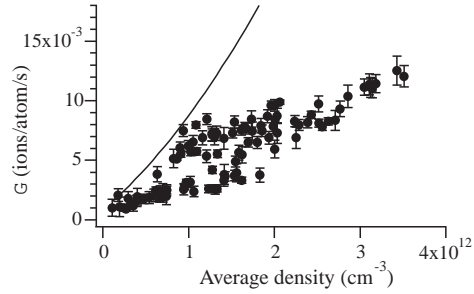}
\end{center}
\caption{\textit{Ion rate per trapped atom ($\Gamma$) in a thermal
cloud versus average density. The solid curve corresponds to the
value of $\beta$ and $L$ deduced from the condensate
measurements.}}\label{toohigh}
\end{figure}

\begin{figure}
\begin{center}
\includegraphics[height=4cm]{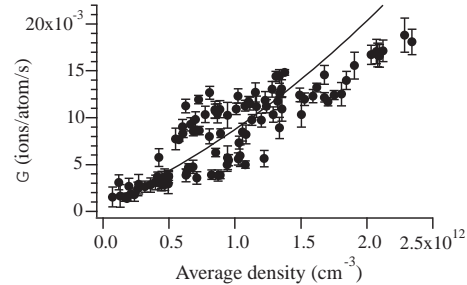}
\end{center}
\caption{\textit{Same as Fig. \ref{toohigh} but assuming a factor
of 1.5 higher detection efficiency of the thermal cloud relative
to the BEC. The data have simply been re-scaled along both axes;
the solid curve is the same as in Fig.
\ref{toohigh}.}}\label{fits}
\end{figure}

As we have shown above, the detection efficiency is expected to be
different for a thermal cloud and we can investigate the effect of
this systematic error on these measurements. We repeat the above
described experiment, this time with a pure thermal cloud. To
begin with, we assume that the detection efficiency is the same
for a BEC and a thermal cloud. We plot the ion rate per atom as a
function of $\langle n \rangle$ in Fig. \ref{toohigh}. We can
extrapolate the data to obtain the vertical intercept, which
corresponds to $1/\tau'$. For densities corresponding to the
moment of formation of BEC, the corresponding ion rate $N/\tau'$
is negligible compared to the total ionization rate, meaning that
we are dominated by 2- and 3-body processes (see Fig. \ref{naiss}
and \ref{Natomes}). To compare with the results obtained for the
BEC, we have also plotted the curve we would expect using the
above values of $\beta$ and $L$. It is clear that the data do not
agree with this curve.
Moreover, no possible pair of $\beta$ and $L$ taken within their
error bars (see \cite{prl}) can transform the curve so that it
agrees with the data. Nor can assuming a different scattering
length. What {\it{can}} make the curve agree with the data is
assuming a different detection efficiency for atoms in the thermal
cloud. If we assume for example that the detection efficiency is a
factor of 1.5 higher for a thermal cloud relative to a BEC (which
is consistent with Fig.\ref{Natomes}), the curve agrees better
with the data as shown in Fig. \ref{fits}.

\begin{figure}
\begin{center}
\includegraphics[height=4cm]{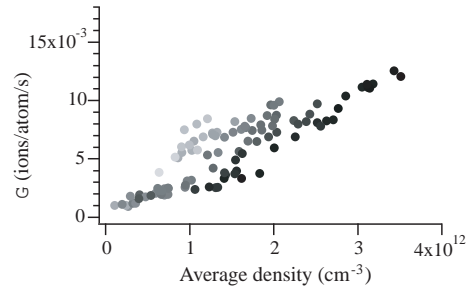}
\end{center}
\caption{\textit{Same data as in Fig. \ref{toohigh} but with the
different temperatures indicated. The light grey corresponds to
the point with higher temperatures (the maximum temperature is 5.5
$\mu K$), the black points with lower temperatures (minimum
temperature 1.8 $\mu K$).}}\label{temp}
\end{figure}

The dispersion of the data points is quite large. This dispersion
can be understood by examining Fig. \ref{temp} in which we have
plotted the same data as in Fig. \ref{toohigh}, but now indicating
the temperature corresponding to each different point on the
graph. There is a clear systematic variation with temperature. One
possible explanation is that the detection efficiency is
temperature dependent. This agrees with the above idea that the
efficiency depends on the spatial extent of the cloud which is
indeed related to the temperature. We do not know the form of the
detection efficiency as a function of temperature, but comparing
these data (indicating that cold atoms are better detected) with
the fact that a thermal cloud is better detected than a BEC, leads
us to conclude that there exists a certain temperature giving a
maximal detection efficiency.
Therefore the correction to the detection efficiency for thermal
atoms is not just a simple factor, but rather a function of
temperature. Without knowing this correction, we cannot use this
method to determine the collision constants for a thermal cloud.
Still, these results are a consistency check on the rate constants
measured using a BEC.

\section{\bf Conclusion}
We have seen that the benefits of ion detection are twofold.
First, the ion rate can be used to select BECs' with very similar
parameters out of a sample with large fluctuations. Second, the
ion rate itself can give information on the condensate on a single
shot basis. Quantitatively, we still have some difficulties
interpreting the data due to systematic errors in the detection
calibration.

One way to overcome this problem is to release the atoms from the
trap by the mean of Raman transitions. It should be possible to
transfer close to $100\%$ of the atoms into the $m=0$ state. This
will eliminate the temperature dependence of the detection
efficiency and allow us to obtain more precise measurements of
$\beta$ and $L$, both for the BEC by improving the value of the
scattering length and for the thermal cloud by making the
detection efficiency temperature independent.

\section{\bf Acknowledgements}

We thank A. Villing and F. Moron for their assistance with the
electronics. This work is supported by the European Union under
grants IST-1999-11055 and HPRN-CT-2000-00125, and by the DGA grant
00.34.025. J.V.G. thanks Funda\c c\~ao para a Ci\^encia e a
Tecnologia and S.S. acknowledges Danish Research Training Council
and Danish Research Agency for financial support.

\appendix

\section{Predictions of the ion rate during the formation of the BEC}
\label{ions}

The 2- and 3-body ion rates ($I_{2b}$ and $I_{3b}$, respectively)
in a sample containing both a BEC and a thermal cloud is given by
\cite{DepletionGora}:
\begin{equation}
I_{2b} = \frac{\beta}{2!} \int d{\bf r}\left[ n_0^2({\bf r}) + 4
n_0({\bf r}) n_{th}({\bf r}) + 2 n_{th}^2({\bf r})
\right]\label{i2b}
\end{equation}
\begin{equation}
I_{3b} = \frac{L}{3!} \int d{\bf r}\left[ n_0^3({\bf r}) + 9
n_0^2({\bf r}) n_{th}({\bf r}) + 18 n_0({\bf r}) n_{th}^2({\bf r})
+ 6 n_{th}^3({\bf r}) \right]\label{i3b}
\end{equation}
where $n_0({\bf r})$ is the local density of the BEC and
$n_{th}({\bf r})$ is the local density of the thermal cloud. Here
we have taken into account the symmetrization factors, but
neglected quantum depletion.

Four parameters are needed to determine the densities of the two
clouds : $N_0$, $\mu$, $N_{th}$ and $T_{th}$. In the Thomas-Fermi
approximation however, the BEC density depends only on $\mu$:
\begin{equation}
n_0({\bf r}) =  \max \left[ 0,\frac{\mu - U({\bf r})}{g} \right]
\end{equation}
with $U({\bf r})$ the harmonic trapping potential and $g=\frac{4
\pi \hbar^2 a}{m}$ the interaction strength. The density of the
thermal cloud depends on two parameters. But, if thermodynamic
equilibrium is reached, taking into account the interactions
between the BEC and the thermal cloud (and neglecting the
interaction energy of the thermal cloud), we can write:
\begin{equation}
n_{th}({\bf r}) = \frac{1}{\lambda_{dB}^3} \, g_{3/2} \left(
\exp^{- \frac{1}{k_B T}( U({\bf r}) + 2g \, n_0({\bf r}) -\mu )}
\right)
\end{equation}
where $\lambda_{dB}$ is the thermal de Broglie wavelength and
$g_{3/2}(x)=\sum\limits_{n=1}^{+\infty} \frac{x^n}{n^{3/2}}$. In
that case, given $\mu$, $n_{th}$ only depends on one additional
parameter.

\subsection{Comparison on the ion rates created by a thermal cloud at $T=T_C$ and a pure BEC}

Before trying to calculate the ion rate for any $T$, which
requires numerical calculation, let us first examine the ion rate
created by a thermal cloud at $T=T_C$ with a number of atoms $N$
and that created by a pure BEC ($T=0$) with a number of atoms
$\eta N$ ($\eta < 1$).

In the case of 2-body collisions, the ratio $R_{2b}$ of the ion
rates created by a pure BEC ($I^{BEC}$) and by a thermal cloud
($I^{th}$) is related to the ratio of the peak densities. For
3-body collisions the ratio ($R_{3b}$) is related to the square of
that ratio. Using the above equations we find:
\begin{equation}
\left( \frac{n_0}{n_{th}} \right) = C_0 \times \eta^{2/5} \times
N^{-1/10} \left( \frac{\overline{\sigma}}{a} \right)^{3/5}
\end{equation}
\begin{equation}
R_{2b} = \frac{I^{BEC}_{2b}}{I^{th}_{2b}} = C_2 \times \eta^{7/5}
\times N^{-1/10} \left( \frac{\overline{\sigma}}{a} \right)^{3/5}
\end{equation}
\begin{equation}
R_{3b} = \frac{I^{BEC}_{3b}}{I^{th}_{3b}} = C_3 \times \eta^{9/7}
\times N^{-2/10} \left( \frac{\overline{\sigma}}{a} \right)^{6/5}
\end{equation}
where $\overline{\sigma}=\sqrt{\frac{\hbar}{m
\overline{\omega}}}$. The numerical factors $C_0 \simeq 0.78$,
$C_2 \simeq 1.05$ and $C_3 \simeq 0.49$ are independent of the
atom considered and only assume that the cloud is trapped in a 3D
harmonic trap. The maximum ratios are reached in the case of no
loss ($\eta = 1$). Using the typical values of our experiment ($a
\simeq 20$~nm, $N \simeq 4. \times 10^5$, and $\overline{\omega}
\simeq 2\pi \times 408$~Hz), we find  $\left( \frac{n_0}{n_{th}}
\right)_{max}\simeq 4$, $(R_{2b})_{max} \simeq 5$ and
$(R_{3b})_{max} \simeq 12$.

If the total number of atoms decreases during the formation of the
BEC, these ratios rapidly fall. For instance, if the number of
atoms decreases by a factor of 3.5 during the last 750~ms of
evaporation as shown in Fig. \ref{Natomes}, we would not have seen
an increase of ionization rate but roughly the same ion rate at
$t=-750$~ms and at $t=0$~ms ! This is an additional evidence of
the difference of neutral atom detection efficiency for a thermal
cloud and BEC (i.e. the total number of atoms decreased by less
than 3.5).

\subsection{Evolution of the ion rate between $T=T_C$ and $T=0$}

Using equation (\ref{i2b}) and  (\ref{i3b}), we have numerically
calculated the ion rates for all temperatures. If the cloud is at
thermodynamic equilibrium all the parameters of the cloud are
deduced from 2 parameters, for example the total number of atoms
and the temperature. To simulate a time evolution of the ion rate
we thus need a model for the variation of these parameters. In
this appendix we will assume a linear evolution of the temperature
between $T=T_C$ and $T=0$ in 0.7 sec. This is of course a
simplification, but given the linearity of the evaporative cooling
ramp, it is a quite good approximation.

\begin{figure}
\begin{center}
\includegraphics[height=4cm]{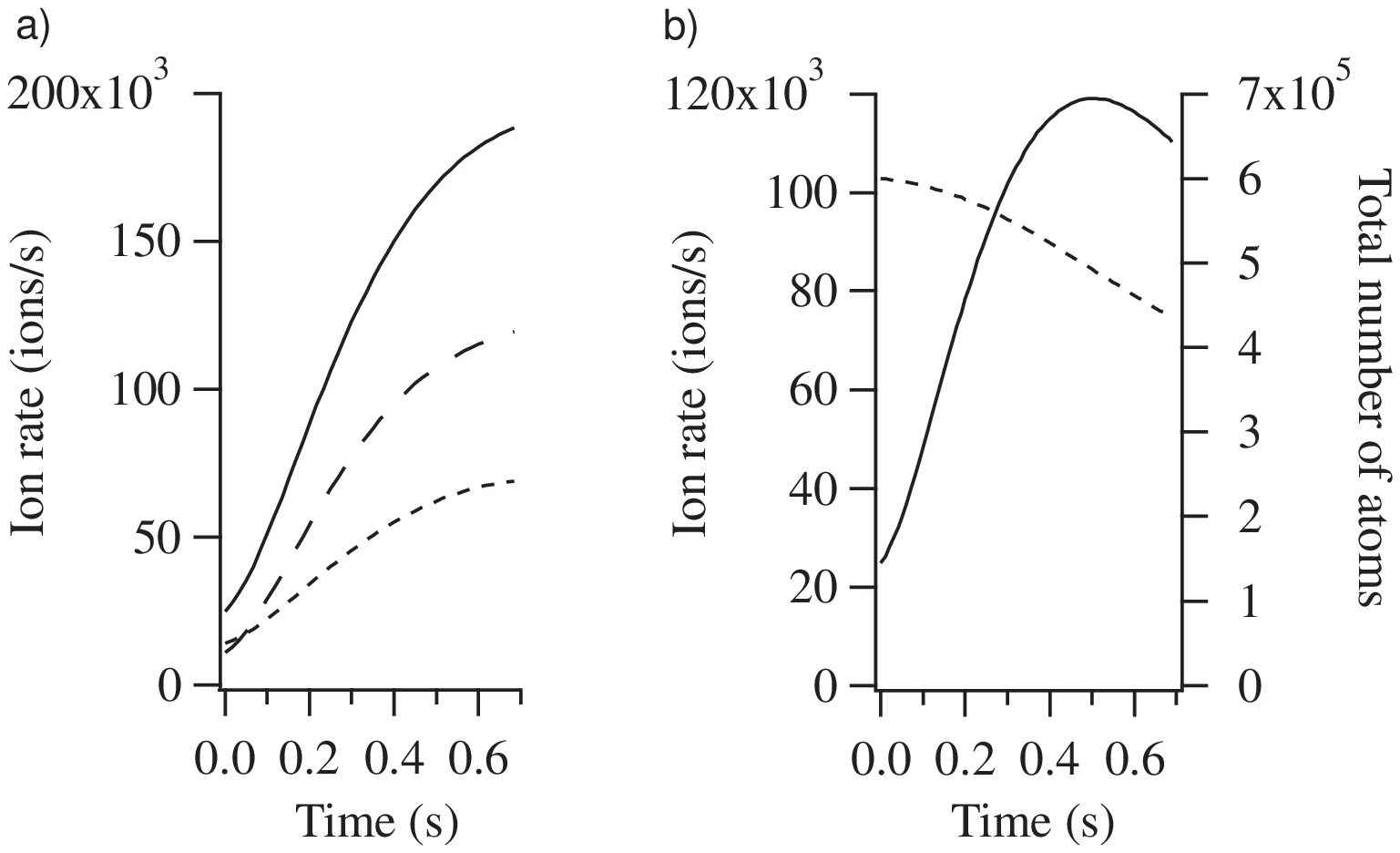}
\end{center}
\caption{Evolution of the ion rate with time. In a) the total
number of atoms is constant. We show the different contributions
to the total ion rate (continuous curve) of the ion rate created
by 2-body collisions (short dashed curve) and 3-body collisions
(long dashed curve.) In b) the total number of atoms (dashed
curve) decrease with the only effect of ionizing collisions. Here
the ion rate exhibits a maximum before the formation of a pure
BEC. For both graphs, the initial number of atoms is $6 \times
10^5$ and a linear evolution of the temperature between $T_C$ and
0 is imposed. The rates have been calculated with the values of
rate constants measured in \cite{prl}.}\label{evol1}
\end{figure}

In Fig. \ref{evol1} a) we show the evolution of the ion rates
assuming a constant total number of atoms. The ion rate increases
monotonically. We also see that the number of ions produced and
thus also the number of lost atoms is not necessarily negligible
compared to the total.

We can attempt to take into account these losses in our model. In
the experiments described in the text, the losses are not only due
to the ionizing collisions but also to the rf-knife. In addition,
losses not only lead to a decrease of the total number of atoms
but also to a change of the temperature because these collisions
change the condensed fraction. Thus, modelling the ion rate can be
quite complicated. Here we wish simply to illustrate the effect of
loss, so we assume that losses are only due to ionizing
collisions, and we will neglect losses due to the rf-knife. Figure
\ref{evol1} b) shows the results. The atom number decreases by
only 30 \% and the ion rate reaches a local maximum before the
formation of the pure BEC, as in our experiment. Extensions of our
model to include losses due to the rf-knife would allow one to
monitor all the parameters of the cloud using the ion signal.

\section{Proposed measurement of rate constants independent of absolute neutral atom detection efficiency}
\label{detection}

We will assume in this section that the absolute ion detection
efficiency is known, and that 2- and 3-body losses are ionizing
collisions \cite{Shlyapnikov:94a}. The idea behind this method is
that two TOF signals separated by a given time can measure the
relative atom loss during this time, while the ion rate can
measure the absolute atom loss. These data allow one to extract
the rate constants without relying on an absolute calibration of
the neutral atom detection efficiency. The method works if the
neutral detection efficiency is unknown, but independent of
temperature. Otherwise, we must also assume that the cloud does
not heat during the measurement or that we know the variation of
detection efficiency with temperature.

To simplify the discussion we will neglect 3-body reactions and
assume that the sample does not heat during the measurement. This
will allow us to derive analytical expressions, but the results
are easily generalized to include heating as well as 3-body
reactions. We can then write the ion rate $I(t)$ as:
\begin{equation}\label{betatau}
I(t)=\frac{\epsilon N(t)}{\tau'}+\frac{\beta
\epsilon}{2V_{eff}}N(t)^{2}
\end{equation}
with $\tau'$ the life time due to ionizing collisions, $N(t)$ the
absolute atom number, $V_{eff}$ defined by $\langle n
\rangle=\frac{N}{V_{eff}}$ and $\epsilon$ the ion detection
efficiency. We write $N_{d}(t)=\alpha N(t)$ where $N_{d}(t)$ is
the detected number of atoms, and $\alpha$ is the neutral atom
detection efficiency. Then
\begin{equation}\label{betatau1}
I(t)=\frac{\epsilon N_{d}(t)}{\alpha\tau'}+\frac{\epsilon
\beta}{\alpha^{2} 2V_{eff}} N_{d}(t)^{2}
\end{equation}
We can also write an equation for the atom number
\begin{equation}\label{betatau2}
\frac{dN(t)}{dt}=-\frac{N(t)}{\tau}-\frac{\beta}{V_{eff}} N(t)^{2}
\end{equation}
with $\tau$ the total life time of the sample that we can measure
independently at lowest density. The solution is:
\begin{equation}\label{betatau3}
\frac{N(t)}{N(t_{0})}=\frac{1}{(1+\frac{\beta}{V_{eff}}
N(t_{0})\tau)e^{(t-t_{0})/\tau}-\frac{\beta}{V_{eff}} N(t_{0})\tau
}
\end{equation}
substituting again $N_{d}(t)=\alpha N(t)$ we have:
\begin{equation}\label{betatau4}
\frac{N_{d}(t)}{N_{d}(t_{0})}=\frac{1}{(1+ \frac{\beta}{\alpha
V_{eff}} N_{d}(t_{0})\tau)e^{(t-t_{0})/\tau}- \frac{\beta}{\alpha
V_{eff}} N_{d}(t_{0})\tau }
\end{equation}
Thus we can measure an initial ion rate and the corresponding
detected atom number $N_{d}(t_{0})$ by a TOF signal, let the
system evolve during a certain time and then again measure the ion
rate and the atom number $N_{d}(t)$. With the evolution of the ion
rate, we can deduce $\epsilon/\alpha\tau'$ and $\frac{\epsilon
\beta}{\alpha^{2} V_{eff}}$ from equation (\ref{betatau}), and
from the evolution of the atom number we can deduce
$\frac{\beta}{\alpha V_{eff}} $ using equation (\ref{betatau4}).
With the value of $V_{eff}$ and $\epsilon$, we can obtain the
value $\beta$. We can also obtain the detection efficiency
$\alpha$.

If we allow for three-body reactions, the method can still be used
but (\ref{betatau3}) is no longer analytical and must be
integrated numerically. If the sample heats during the
measurement, we only have to recalculate the volume $V_{eff}$ for
each TOF measurement.

The reason why we have yet not been able to apply this method is
as indicated above that the sample is heating so that the
detection efficiency changes during the measurement. As we have
not been able to measure the temperature dependence of $\alpha(T)$
the above equations cannot be solved. We hope to render the
detection efficiency temperature independent in the near future by
using Raman transitions as mentioned in the conclusion.

\smallskip
\smallskip
\smallskip



\begin{thebibliography}{9}
\bibitem{notedefcoef}
Collision rate constants are sometimes defined directly for a BEC
($\beta'=\beta/2$ and $L'=L/6$).
\end{thebibliography}
\end{document}